\documentclass[twocolumn,showpacs,amsmath,amssymb]{revtex4}

\usepackage{graphicx}
\usepackage{dcolumn}
\usepackage{bm}

\begin{document}


\title{Differential cross section and photon beam asymmetry for 
the $\vec{\gamma} n$ $\rightarrow$ $K^{+}\Sigma^{-}$ reaction 
at $E_{\gamma}$=1.5-2.4 GeV}

\author{H.~Kohri$^{1}$, 
D.S.~Ahn$^{1,2}$, 
J.K.~Ahn$^{2}$, 
H.~Akimune$^{3}$, 
Y.~Asano$^{4}$,
W.C.~Chang$^{5}$, 
S.~Date'$^{6}$, 
H.~Ejiri$^{1,6}$, 
S.~Fukui$^{7}$, 
H.~Fujimura$^{8,9}$, 
M.~Fujiwara$^{1,4}$, 
S.~Hasegawa$^{1}$, 
K.~Hicks$^{10}$, 
T.~Hotta$^{1}$, 
K.~Imai$^{9}$, 
T.~Ishikawa$^{11}$, 
T.~Iwata$^{12}$, 
H.~Kawai$^{13}$, 
Z.Y.~Kim$^{8}$,
K.~Kino$^{1,a}$, 
N.~Kumagai$^{6}$, 
S.~Makino$^{14}$, 
T.~Mart$^{15}$,
T.~Matsuda$^{16}$, 
T.~Matsumura$^{1,4,b}$, 
N.~Matsuoka$^{1}$, 
T.~Mibe$^{1,4,10}$, 
M.~Miyabe$^{9}$,
Y.~Miyachi$^{17}$, 
M.~Morita$^{1}$, 
N.~Muramatsu$^{4,1}$, 
T.~Nakano$^{1}$, 
M.~Niiyama$^{9}$, 
M.~Nomachi$^{18}$, 
Y.~Ohashi$^{6}$,  
H.~Ohkuma$^{6}$, 
T.~Ooba$^{13}$, 
D.S.~Oshuev$^{5,c}$, 
C.~Rangacharyulu$^{19}$, 
A.~Sakaguchi$^{18}$, 
T.~Sasaki$^{9}$, 
P.M.~Shagin$^{20}$, 
Y.~Shiino$^{13}$, 
A.~Shimizu$^{1}$, 
H.~Shimizu$^{11}$, 
Y.~Sugaya$^{18}$, 
M.~Sumihama$^{18,4}$, 
Y.~Toi$^{16}$, 
H.~Toyokawa$^{6}$, 
A.~Wakai$^{21}$, 
C.W.~Wang$^{5}$, 
S.C.~Wang$^{5}$, 
K.~Yonehara$^{3,d}$, 
T.~Yorita$^{1,6}$, 
M.~Yoshimura$^{22}$, 
M.~Yosoi$^{9,1}$, 
and R.G.T.~Zegers$^{23}$\\
(LEPS Collaboration)\\}

\affiliation{$^{1}$Research Center for Nuclear Physics, Osaka 
University, Ibaraki, Osaka 567-0047, Japan}
\affiliation{$^{2}$Department of Physics, Pusan National University, 
Busan 609-735, Korea}
\affiliation{$^{3}$Department of Physics, Konan University, 
Kobe, Hyogo 658-8501, Japan}
\affiliation{$^{4}$Kansai Photon Science Institute, Japan Atomic Energy 
Agency, Kizu, Kyoto 619-0215, Japan}
\affiliation{$^{5}$Institute of Physics, Academia Sinica, Taipei, Taiwan 11529, 
Republic of China}
\affiliation{$^{6}$Japan Synchrotron Radiation Research Institute, 
Mikazuki, Hyogo 679-5198, Japan}
\affiliation{$^{7}$Department of Physics and Astrophysics, Nagoya University, 
Nagoya, Aichi 464-8602, Japan}
\affiliation{$^{8}$School of Physics, Seoul National University, Seoul, 
151-747, Korea}
\affiliation{$^{9}$Department of Physics, Kyoto University, 
Kyoto 606-8502, Japan} 
\affiliation{$^{10}$Department of Physics And Astronomy, Ohio University, 
Athens, Ohio 45701, USA}
\affiliation{$^{11}$Laboratory of Nuclear Science, Tohoku University, 
Sendai, Miyagi 982-0826, Japan}
\affiliation{$^{12}$Department of Physics, Yamagata University, 
Yamagata 990-8560, Japan}
\affiliation{$^{13}$Department of Physics, Chiba University, 
Chiba 263-8522, Japan}
\affiliation{$^{14}$Wakayama Medical College, Wakayama, 
Wakayama 641-8509, Japan}
\affiliation{$^{15}$Departemen Fisika, FMIPA, Universitas Indonesia, 
Depok 16424, Indonesia}
\affiliation{$^{16}$Department of Applied Physics, Miyazaki University, 
Miyazaki 889-2192, Japan}
\affiliation{$^{17}$Department of Physics, Tokyo Institute of Technology, 
Tokyo 152-8551, Japan} 
\affiliation{$^{18}$Department of Physics, Osaka University, Toyonaka, 
Osaka 560-0043, Japan}
\affiliation{$^{19}$Department of Physics, 
University of Saskatchewan, Saskatoon, Saskatchewan, Canada} 
\affiliation{$^{20}$School of Physics and Astronomy, University of Minnesota, 
Minneapolis, Minnesota 55455, USA}
\affiliation{$^{21}$Akita Research Institute of Brain and Blood Vessels, 
Akita 010-0874, Japan}
\affiliation{$^{22}$Institute for Protein Research, Osaka University, 
Suita, Osaka 565-0871, Japan}
\affiliation{$^{23}$Department of Physics and Astronomy, 
Michigan State University, East Lansing, MI 48824-1321, USA}

\date{\today}
\begin{abstract}
Differential cross sections and photon beam asymmetries 
have been measured 
for the $\vec{\gamma} n$ $\rightarrow$ $K^{+}\Sigma^{-}$ 
and $\vec{\gamma} p$ $\rightarrow$ $K^{+}\Sigma^{0}$ reactions 
separately using liquid deuterium and hydrogen targets with 
incident linearly polarized photon beams of 
$E_{\gamma}$=1.5-2.4 GeV at 0.6$<\cos\Theta^{K}_{\rm cm}<$1. 
The cross section ratio of 
$\sigma_{K^{+}\Sigma^{-}}$/$\sigma_{K^{+}\Sigma^{0}}$, expected 
to be 2 on the basis of the isospin 1/2 exchange, 
is found to be close to 1. 
For the $K^{+}\Sigma^{-}$ reaction, large positive asymmetries 
are observed indicating the dominance of $K^{*}$-exchange. 
The large difference between the asymmetries 
for the $K^{+}\Sigma^{-}$ and $K^{+}\Sigma^{0}$ reactions can 
not be explained by simple theoretical considerations 
based on Regge model calculations. 
\end{abstract}

\maketitle

The reaction mechanism of strangeness photoproduction 
is important to study in order to understand the 
role of nucleon resonances, the hyperon resonances, 
and their decay branching ratios.
By comparing data and theoretical models, we gain a 
deeper understanding of the underlying dynamics, 
which are often described in terms of constituent 
quarks as the effective degrees of freedom.
For example, many hadron decays are described successfully 
in terms of the $^{3}P_{0}$ model \cite{LeYaouanc}, where 
the strange quark-antiquark ($s\bar{s}$) pair has the quantum 
numbers of the vacuum, produced with its spins aligned. 
However, this model is cast into doubt by 
a recent JLAB/CLAS experiment \cite{Carman}, where 
the observation of the transferred polarization 
for the $\vec{e}p$ $\rightarrow$ $e'K^{+}\vec{\Lambda}$ 
reaction showed that the $s\bar{s}$ pair is predominantly 
produced with its spins antialigned. 
While this result may be particular to the reaction 
studied, and not necessarily the general case, it implies that 
our understanding of the mechanisms of the baryon decays and 
$s\bar{s}$ pair production is incomplete.  
The measurement of polarization observables, in addition to the cross 
section, provides more information to reveal the processes 
given above. 
Herein we present the photon beam asymmetry for production 
of the $\Sigma$ hyperon, which provides new polarization data.

Theoretically, kaon photoproduction is 
described in terms of hadron exchanges, such as 
$N$, $N^{*}$, and $\Delta^{*}$ in the $s$-channel, 
hyperon and excited hyperon in the $u$-channel, 
and $K$ and $K^{*}$ in the $t$-channel. 
The contribution from meson exchanges in the $t$-channel is 
expected to be large at forward angles. 
The photon beam asymmetry has a unique feature: 
at small $|t|$ and at high energies \cite{Stichel,Guidal} 
its value is $+$1 or $-$1 if the $K^{*}$ or $K$ meson 
is exchanged in the $t$-channel, respectively. 
Thus, nucleon resonance contributions appear as 
modulations in photon beam asymmetries 
at the photon energy corresponding to a given 
resonance mass.
The measured asymmetries for the 
$\vec{\gamma} p$ $\rightarrow$ $K^{+}\Lambda, K^{+}\Sigma^{0}$ 
reactions at forward angles were close to $+$1 at high 
energies (above the resonance region) of $E_{\gamma}$=5-16 GeV 
\cite{Quinn}, whereas they were 
significantly smaller for the first reaction at photon energies of 
$E_{\gamma}$=1.5-2.4 GeV \cite{Zegers}. 

Experimental information on the $N^{*}$ and $\Delta^{*}$ 
resonances has been obtained primarily in studies of their 
pionic decays.
Constituent quark models predict more nucleon resonances 
than have been observed experimentally. 
These unobserved nucleon resonances are called 
'missing resonances'. 
Quark model studies based on the $^{3}P_{0}$ model 
suggest that these resonances can couple to strangeness 
channels, such as $K\Lambda$ and $K\Sigma$ \cite{Capstick}. 
According to simple isospin arguments for the $K\Sigma$ 
channels, $N^{*}$ resonances couple strongly to 
$K^{0}\Sigma^{+}$ and $K^{+}\Sigma^{-}$ channels, 
while $\Delta^{*}$ resonances couple strongly to 
$K^{+}\Sigma^{0}$ and $K^{0}\Sigma^{0}$ channels. 
Therefore, the comparison between $K^{+}\Sigma^{-}$ and 
$K^{+}\Sigma^{0}$ is an important tool in identifying 
contributions from $N^{*}$ or $\Delta^{*}$ resonances. 

Five nucleon resonances, $S_{11}$(1650), $P_{11}$(1710), 
$P_{13}$(1720), $S_{31}$(1900), and $P_{31}$(1910) 
are well known in kaon photoproduction. 
Recent data for the $K^{+}\Lambda$ channel clarified the 
existence of a new $D_{13}$ nucleon resonance
at around 1900 MeV \cite{Tran,Glander,Mcnabb,Sumihama}. 
This demonstrates the utility of approaching the missing 
resonance problem using strange baryon production data.

In the past, experimental data were limited 
to $K^{+}\Lambda$ and $K^{+}\Sigma^{0}$ productions off the proton 
\cite{Quinn,Boyarski1,Boyarski2,Zegers,Tran,Glander,Mcnabb,Sumihama,Bradford}. 
Recently, an experimental result for $K^{0}\Sigma^{+}$ 
off the proton was reported \cite{Lawall}. 
However, no experimental result has yet been published for 
kaon photoproduction off the neutron although there exist 
$d$($\gamma$,$K^{+}$) measurements from the 1970's where 
the $K^{+}\Sigma^{-}$ is not separated from 
the $K^{+}\Sigma^{0}$ \cite{Boyarski2,Quinn}. 
In this Letter, we present, for the first time,
differential cross sections and photon beam asymmetries for the 
$\vec{\gamma} n$ $\rightarrow$ $K^{+}\Sigma^{-}$ reaction and 
compare with the 
$\vec{\gamma} p$ $\rightarrow$ $K^{+}\Sigma^{0}$ reaction. 


The experiment was carried out using the laser-electron 
photon facility at SPring-8 (LEPS) \cite{Nakano}. 
Linearly polarized photons were produced by backward Compton 
scattering of an ultra-violet Ar laser from 8 GeV electrons. 
The energy range of tagged photons was 1.5-2.4 GeV, and 
the photon polarization was typically 90\% at 2.4 GeV. 
The experimental setup was described in detail 
elsewhere \cite{Sumihama}. 
Liquid hydrogen (LH$_{2}$) and deuterium (LD$_{2}$) targets with 
an effective length of 16 cm were employed.


Charged particles were detected at forward angles. 
The particle identification was achieved by using the 
time-of-flight and momentum information. 
The events of $K^{+}$ mesons were identified within 3$\sigma$, 
where $\sigma$ is the momentum dependent mass resolution. 
The contamination of $\pi^{+}$ mesons in the $K^{+}$ sample 
was smaller than 5\%. 
The contribution of the target windows and the plastic scintillator 
behind the target was smaller than 4\%. 

\begin{figure}
\begin{center}
\includegraphics[width=7.8cm,height=6.3cm]{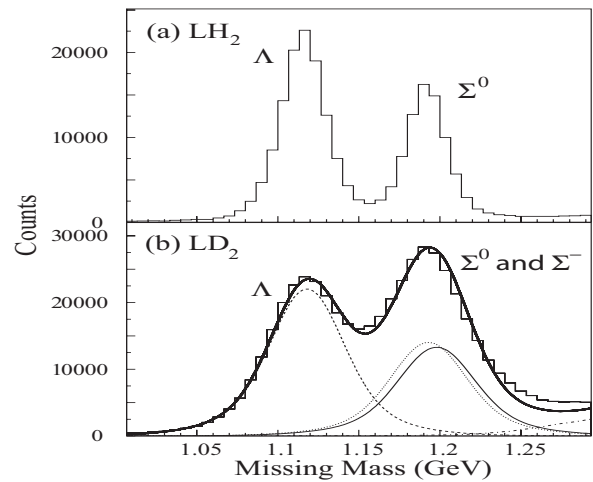}%
\caption{\label{fig_1} Missing mass spectra obtained for 
$K^{+}$ photoproduction off LH$_{2}$ (a) and LD$_{2}$ (b) targets.  
The dashed, dotted, and solid curves correspond to 
$\Lambda$(1116), $\Sigma^{0}$(1193), and $\Sigma^{-}$(1197) 
production, respectively. 
The dot-dashed curve is the estimated background. 
The thick solid curve is the sum of all contributions. 
The thick solid curve deviates from the data histogram at 
$MM_{\gamma K^{+}}>$1.24 GeV because the $\pi^{+}$ 
contamination has not been subtracted from the data histogram. }
\label{fig:missall} 
\end{center}
\end{figure}

Figure \ref{fig:missall} shows the missing mass 
($MM_{\gamma K^{+}}$) spectra for the $p$($\vec{\gamma}$,$K^{+}$)$X$ 
(LH$_{2}$) and $N$($\vec{\gamma}$,$K^{+}$)$X$ (LD$_{2}$) 
reactions for $E_{\gamma}$=1.5-2.4 GeV and 
0.6$<\cos\Theta_{\rm cm}<$1. 
For the LD$_{2}$ data the target was assumed to have 
a mass of ($M_{p}$+$M_{n}$)/2 and zero momentum for the 
$MM_{\gamma K^{+}}$ calculation. 
For the LD$_{2}$ data the peak widths are wider than 
those for the LH$_{2}$ data due to Fermi motion. 
$\Lambda$ and $\Sigma^{0}$ particles are produced on the 
proton, while the $\Sigma^{-}$ particle is produced 
on the neutron.
Therefore, the ratio 
$N$($\Sigma$)/$N$($\Lambda$) in the LD$_{2}$ data is larger 
than the ratio $N$($\Sigma^{0}$)/$N$($\Lambda$) 
in the LH$_{2}$ data. 
The cross section for hadron photoproduction on a bound 
proton in deuteron is almost the same as for a free proton 
if the nuclear effects, 
such as final-state interaction and shadowing effects, 
are small. 
In this analysis, the ratio $N$($\Sigma^{0}$)/$N$($\Lambda$) 
for the LD$_{2}$ data was assumed to be the same as for 
the LH$_{2}$ data, and nuclear effects were evaluated 
as systematic errors. 
The $K^{+}\Sigma^{-}$ cross sections can be 
obtained from the difference between the production yield 
ratios of 
$N$($\Sigma$)/$N$($\Lambda$) in the LD$_{2}$ data 
and $N$($\Sigma^{0}$)/$N$($\Lambda$) in the LH$_{2}$ data.  

The $K^{+}$ angular region in the center-of-mass (cm) system 
was divided into 4 bins and the photon energy region of 1.5-2.4 GeV 
was divided into 18 bins. 
The production yields of $\Lambda$, $\Sigma^{0}$, $\Sigma^{-}$, 
and the background under the $\Sigma$ peak were obtained 
by a fit to the missing mass spectrum with six free parameters. 
The peak shape of each hyperon was estimated by assuming a 
quasi-free reaction process in the GEANT simulation, where 
the Paris potential \cite{Lacombe} was used to generate the initial 
nuclear momentum distribution in deuterium. 
The peak shape was reproduced by the sum of two Gaussians 
having different widths and amplitudes, and was fixed in the fit. 
Two free parameters were used to scale the heights of 
the $\Lambda$ and $\Sigma^{-}$ peaks. 
The production yield ratio of $N$($\Sigma^{0}$)/$N$($\Lambda$) 
gave the height of the $\Sigma^{0}$ peak. 
The peak position of $\Lambda$ was a free parameter, and 
the $\Sigma^{0}$ and $\Sigma^{-}$ peaks were 
placed at 0.077 GeV and 0.082 GeV higher than 
the $\Lambda$ peak, respectively. 
The main background under the $\Sigma$ peak was considered to 
be due to $\pi \Lambda$, $\pi \Sigma$, and the tail of 
$\Lambda^{*}$(1405) and $\Sigma^{*}$(1385) events. 
The distribution shape of each reaction was estimated by 
the GEANT simulation. 
Two free parameters were scale factors for the heights of 
the $\pi \Lambda$ and $\pi \Sigma$ events, and one free parameter 
was used for the tail of 
the $\Lambda^{*}$(1405) and $\Sigma^{*}$(1385) events. 

As a result of the fit, the production yield 
ratio $N$($\Sigma^{-}$)/$N$($\Sigma^{0}$) was obtained. 
The $K^{+}\Sigma^{0}$ cross section was 
obtained from the LH$_{2}$ data using the same method 
as Ref.\cite{Sumihama}.  
The $K^{+}\Sigma^{-}$ cross section was calculated by using 
the ratio, 
\begin{equation}
\frac{d\sigma_{\Sigma^{-}}}{d\cos\Theta_{\rm cm}} = 
\frac{d\sigma_{\Sigma^{0}}}{d\cos\Theta_{\rm cm}} \times 
\frac{N(\Sigma^{-})}{N(\Sigma^{0})}. 
\end{equation} 
Since the masses of $\Sigma^{-}$ and $\Sigma^{0}$ are almost the 
same, acceptance corrections are negligible. 


Most of the nuclear effects were small and canceled by taking 
the yield ratios, 
$N$($\Sigma^{0}$)/$N$($\Lambda$) and 
$N$($\Sigma^{-}$)/$N$($\Sigma^{0}$), in the analysis 
because the total cross sections for $\gamma p$ and $K^{+}p$ 
are similar to those for $\gamma n$ and $K^{+}n$, respectively. 
However, differences among $\Lambda n$, $\Sigma^{0} n$, and 
$\Sigma^{-} p$ final-state interactions are not 
negligible \cite{Yamamura}. 
Final-state interactions considered by Yamamura $et$ $al$. 
\cite{Yamamura} using the NSC97f hyperon-nucleon force 
were incorporated in our data analyses 
as systematic errors. 

\begin{figure}
\begin{center}
\includegraphics[width=7.8 cm,height=9.5cm]{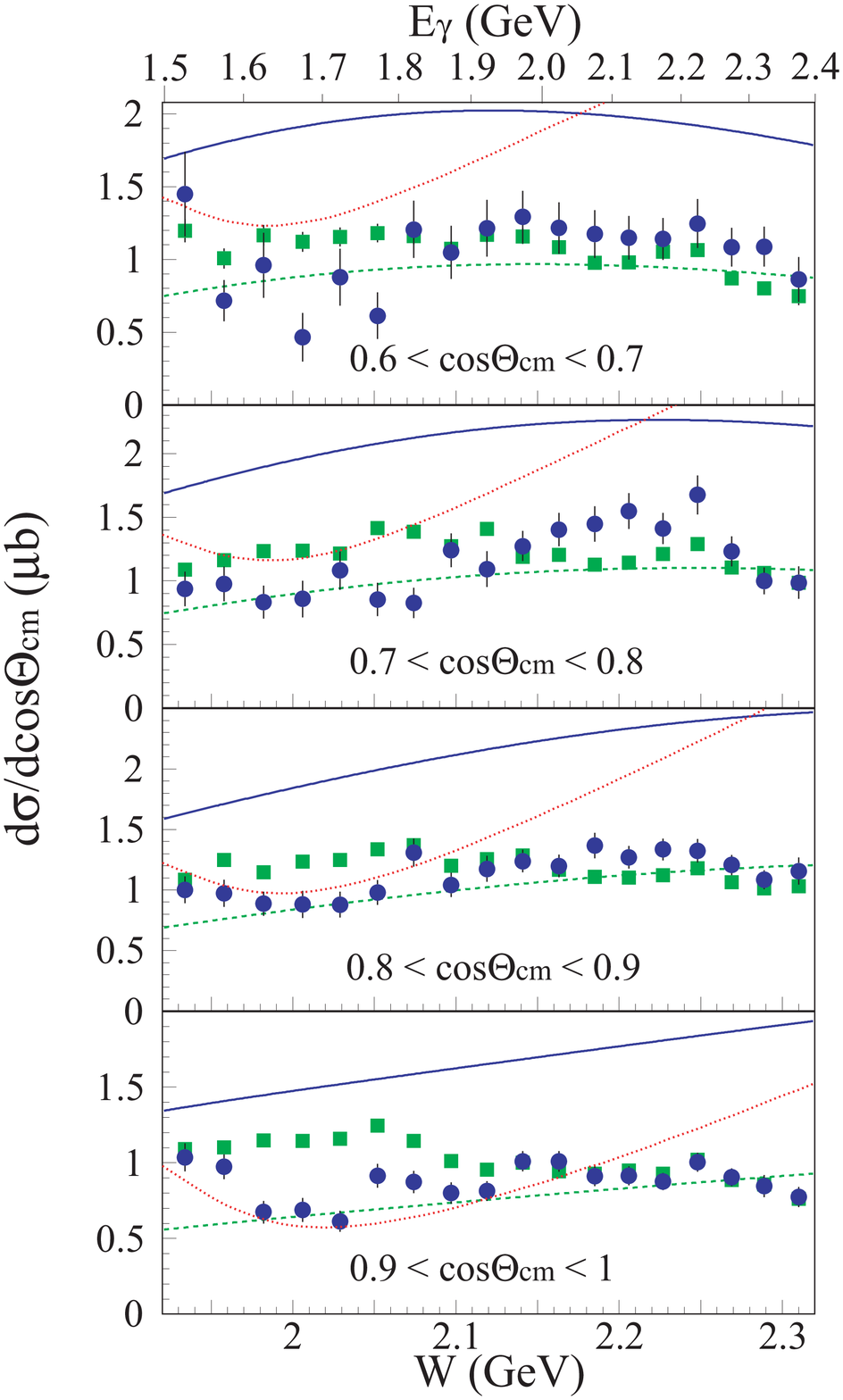}%
\caption{\small Differential cross sections 
for $\vec{\gamma} n$ $\rightarrow$ 
$K^{+}\Sigma^{-}$ (circle) and $\vec{\gamma} p$ $\rightarrow$ 
$K^{+}\Sigma^{0}$ (square). 
Only statistical errors are shown. 
The solid and dashed curves are the Regge model 
calculations \cite{Regge,Guidal} for the $K^{+}\Sigma^{-}$ 
and $K^{+}\Sigma^{0}$, respectively. 
The dotted curve is the KaonMAID model calculations \cite{Kaonmaid} for 
the $K^{+}\Sigma^{-}$. }
\label{fig:cross}
\end{center}
\end{figure}

In Fig.~\ref{fig:cross} the differential cross sections 
for $K^{+}\Sigma^{-}$ and $K^{+}\Sigma^{0}$ are shown. 
The effect of the hyperon-nucleon final-state 
interactions is estimated to be 16\% of the $K^{+}\Sigma^{-}$ 
cross section. 
The effect of the internal two-step process mediated by 
a $\pi$ meson \cite{Salam} is negligible at 0.8$<\cos\Theta_{\rm cm}$, 
and 6\% and 15\% at 0.7$<\cos\Theta_{\rm cm}<$0.8 and 
0.6$<\cos\Theta_{\rm cm}<$0.7, respectively, for the $K^{+}\Sigma^{-}$. 
The Fermi motion effect of the target nucleons is 8\% 
for $K^{+}\Sigma^{-}$. 
Systematic uncertainties of target thickness and 
photon flux are 1\% and 3\%, respectively, for both reactions. 
The effect of the contaminations in the selection of 
$K^{+}$ from the targets was subtracted. 

In the center-of-mass energy ($W$=$\sqrt{s}$) region above 2.1 GeV 
at 0.8$<\cos\Theta_{\rm cm}$, the cross 
sections for the two reactions are similar. 
This result is inconsistent with a model based on 
the exchange of pure isospin 1/2, dominantly due to 
$K$ and $K^{*}$ mesons, where the cross section 
ratio of $\sigma_{K^{+}\Sigma^{-}}$/$\sigma_{K^{+}\Sigma^{0}}$ 
is expected to be 2. 
Regge model calculations \cite{Regge,Guidal} agree with the data for 
$K^{+}\Sigma^{0}$, but largely overestimate the 
$K^{+}\Sigma^{-}$ data. 
At high energies, the KaonMAID model \cite{Kaonmaid} 
also overestimates the data for $K^{+}\Sigma^{-}$. 

Two possibilities are considered to explain the reason 
why the experimental ratio is so different from the 
theoretical expectation. 
One possibility is the contribution from 
$\Delta^{*}$ or $N^{*}$ resonances which may increase the 
cross sections for the $K^{+}\Sigma^{0}$. 
The contribution from the $\Delta^{*}$ resonances may 
reduce the cross sections for the $K^{+}\Sigma^{-}$ by 
destructive interference \cite{Boyarski2}. 
A second possibility is that the $u$-channel $\Lambda$ or 
$\Lambda^{*}$ exchange, which is absent in the $K^{+}\Sigma^{-}$ 
channel, may contribute strongly to the $K^{+}\Sigma^{0}$ channel. 
Since the $N^{*}$ resonances couple weakly to the $K^{+}\Sigma^{0}$ 
channel, and the $u$-channel contribution must be very small at 
forward angles, the $\Delta^{*}$ contribution is the most 
reasonable explanation for the similar cross sections. 
If the isospin 3/2 amplitude is 10\% of the 
isospin 1/2 amplitudes, dominantly due to 
$K$ and $K^{*}$ exchanges, in the $K^{+}\Sigma$ cross 
sections described in Ref.\cite{Boyarski2}, the similarity of 
the $K^{+}\Sigma^{-}$ and $K^{+}\Sigma^{0}$
cross sections can be explained. 


By using vertically and horizontally polarized photon 
beams, the photon beam asymmetry can be measured 
without any correction for the spectrometer 
acceptance \cite{Zegers,Sumihama}. 
The asymmetry ($\Sigma$) is given as follows:
\begin{equation}
P_{\gamma}\Sigma \cos2\Phi = \frac{N_{v}-N_{h}}{N_{v}+N_{h}}, \\
\end{equation}
where $N_{v}$ and $N_{h}$ are the $K^{+}$ photoproduction yields 
with the vertically and horizontally polarized photons, respectively. 
$P_{\gamma}$ is the polarization degree of the photon beam, and 
$\Phi$ is the $K^{+}$ azimuthal angle defined by the angle between 
the reaction plane and the horizontal plane. 
The $K^{+}$ angular region in the cm system was divided 
into 4 bins and the photon energy region of 1.5-2.4 GeV was 
divided into 9 bins. 
In order to obtain the asymmetry for the 
$K^{+}\Sigma^{-}$ separately, the effects of 
background events were subtracted from the asymmetry for 
events selected by the condition 
$|M_{\Sigma^{-}}-MM_{\gamma K^{+}}|<$ $\sigma_{\Sigma^{-}}$, 
where $\sigma_{\Sigma^{-}}$ is the width of the $\Sigma^{-}$ peak. 
The asymmetries for the $K^{+}\Lambda$ and $K^{+}\Sigma^{0}$ 
were obtained from the LH$_{2}$ data using the same method 
as Ref.\cite{Sumihama}. 
The asymmetry for the $K^{+}\pi\Lambda$, $K^{+}\pi\Sigma$, 
and the tail of $K^{+}\Lambda^{*}(1405)$ and 
$K^{+}\Sigma^{*}(1385)$ was obtained from 
events selected by the condition 
1.25$<MM_{\gamma K^{+}}<$1.30 GeV by 
subtracting the effect of the $\Sigma$ peak tail. 
The effect of the contaminations in the selection of 
$K^{+}$ from the targets was subtracted. 

\begin{figure}
\begin{center}
\includegraphics[width=7.8cm,height=9.5cm]{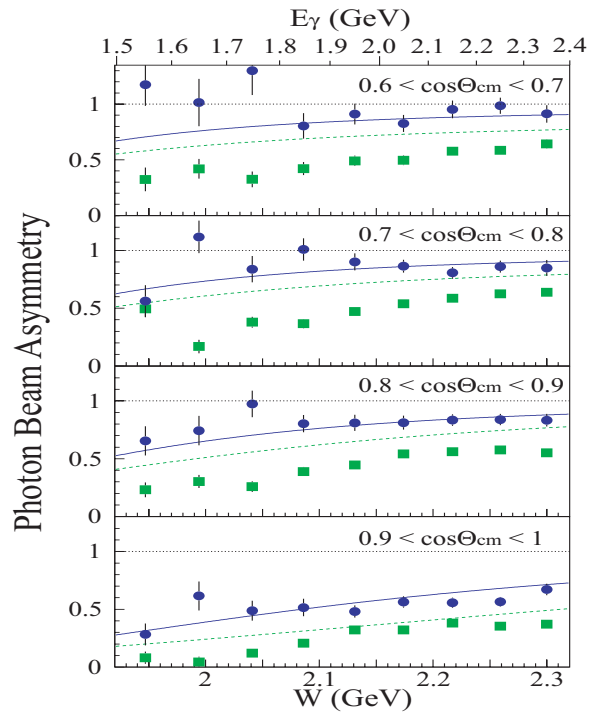}%
\caption{\small Photon beam asymmetries for 
$\vec{\gamma} n$ $\rightarrow$ $K^{+}\Sigma^{-}$ (circle) and 
$\vec{\gamma} p$ $\rightarrow$ $K^{+}\Sigma^{0}$ (square). 
The solid and dashed curves are the Regge model 
calculations \cite{Regge,Guidal} for the $K^{+}\Sigma^{-}$ 
and $K^{+}\Sigma^{0}$, respectively. }
\label{fig:basigmam}
\end{center}
\end{figure}

In Fig.~\ref{fig:basigmam} the photon beam 
asymmetries for $K^{+}\Sigma^{-}$ and $K^{+}\Sigma^{0}$ are shown. 
The effect of the final-state interactions \cite{Yamamura} 
is estimated to be smaller than $\delta\Sigma$=0.1 for $K^{+}\Sigma^{-}$. 
The effect of the internal two-step process mediated by a 
$\pi$ meson \cite{Salam} is negligible at 0.8$<\cos\Theta_{\rm cm}$, 
and $\delta\Sigma$=0.09 and $\delta\Sigma$=0.13 at 
0.7$<\cos\Theta_{\rm cm}<$0.8 and 
0.6$<\cos\Theta_{\rm cm}<$0.7, respectively, for the $K^{+}\Sigma^{-}$. 
The effect due to the Fermi motion of target nucleons is smaller 
than $\delta\Sigma$=0.13 for the $K^{+}\Sigma^{-}$. 
The systematic uncertainty of the measurement of the laser 
polarization is $\delta\Sigma$=0.02 for both reactions. 
The present data also show 
reasonable consistency between the asymmetries 
for the $K^{+}\Lambda$ in the LD$_{2}$ and LH$_{2}$ data. 

For $K^{+}\Sigma^{-}$, the asymmetries are positive and are 
larger than those for $K^{+}\Sigma^{0}$. 
The asymmetries close to $+$1 at $\cos\Theta_{\rm cm}<$0.9 
indicate the dominance of the $K^{*}$ exchange in the $t$-channel. 
The asymmetries are small at 0.9$<\cos\Theta_{\rm cm}$ 
because the asymmetries go to zero at $\cos\Theta_{\rm cm}$=1. 
It is quite interesting that the asymmetries for the 
$K^{+}\Sigma^{0}$ gradually increase with increasing 
center-of-mass energy, while the energy dependence of the asymmetries 
for $K^{+}\Sigma^{-}$ is small at $W>$2 GeV. 

The KaonMAID model \cite{Kaonmaid}, which is not 
shown in the figure, predicts negative asymmetries for 
the $K^{+}\Sigma^{-}$. 
The Regge model calculations \cite{Regge,Guidal} overestimate the data 
for the $K^{+}\Sigma^{0}$, while the calculations agree with 
the data for the $K^{+}\Sigma^{-}$. 
This agreement suggests that an additional contribution, which is 
not included in the calculations, is small in the 
$K^{+}\Sigma^{-}$ channel. As shown above, 
contributions from $\Delta^{*}$ resonances could explain 
the $K^{+}\Sigma^{0}$ data \cite{Kaonmaid}. 
However, this $\Delta^{*}$ contribution reduces the $K^{+}\Sigma^{-}$ 
asymmetries and fails to explain the polarization data. 
One may speculate that the difference between theoretical and 
experimental asymmetries for the $K^{+}\Sigma^{0}$ is, at least 
in part, due to contributions from $u$-channel $\Lambda$ and 
$\Lambda^{*}$ exchanges and $s$-channel $N^{*}$ resonances 
which have much stronger coupling to $\gamma p$ than 
to $\gamma n$. 


In summary, we have measured differential cross sections and photon 
beam asymmetries for $\vec{\gamma} n$ $\rightarrow$ 
$K^{+}\Sigma^{-}$ and $\vec{\gamma} p$ $\rightarrow$ 
$K^{+}\Sigma^{0}$ at $E_{\gamma}$=1.5-2.4 GeV. 
The cross sections for $K^{+}\Sigma^{-}$ are similar to 
those for $K^{+}\Sigma^{0}$ at high energies, 
which is inconsistent with the exchange of pure isospin 1/2. 
Large asymmetries close to $+$1 for the $K^{+}\Sigma^{-}$ 
are found, indicating the dominance of $K^{*}$-exchange 
in the $t$-channel. 
A large difference between the asymmetries for the $K^{+}\Sigma^{-}$ 
and $K^{+}\Sigma^{0}$ can not be explained by simple theoretical 
considerations based on Regge model calculations. 
In the energy region of a few GeV, there is no theoretical 
calculation which describes the strangeness photoproduction well. 
The present result may imply the existence of a hidden reaction 
mechanism, and will provide constrains in the model calculations 
with the aim to advance our understanding of the $s\bar{s}$ 
pair production mechanisms. 

The authors thank the SPring-8 staff for supporting the 
experiment. 
We thank A.I. Titov and B.A. Mecking for fruitful discussions. 
This research was supported in part by the Ministry of 
Education, Science, Sports and Culture of Japan, by 
the National Science Council of Republic of China (Taiwan), 
and by National Science Foundation (USA). 

\begin{small}
\hspace*{-0.4cm}$^{a}$Present address: Faculty and Graduate School of 
Engineering, Hokkaido University, Sapporo 060-8628, Japan. \\
$^{b}$Present address: Department of Applied Physics, National 
Defense Academy, Yokosuka 239-8686, Japan. \\
$^{c}$Present address: Nuclear Physics Institute, Moscow State 
University, Moscow, 119899, Russia. \\
$^{d}$Present address: Illinois Institute of Technology, 
Chicago, IL 60616, USA.\\
\end{small}

\end{document}